\documentclass[conference]{IEEEtran}

\ifCLASSINFOpdf

\else

\fi

\usepackage{cite}

\usepackage{graphicx}
\usepackage{graphics}
\usepackage{epsfig}
\usepackage{epstopdf}

\usepackage{amssymb}
\usepackage[cmex10]{amsmath}
\usepackage{amsthm}
\usepackage{mathrsfs}

\usepackage{longtable}

\usepackage{ifthen,xkeyval,xfor,amsgen}
\usepackage{multicol}
\usepackage{float}
\usepackage{lipsum}
\usepackage[utf8]{inputenc}
\usepackage[english]{babel}

\usepackage{algorithm}
\usepackage{algpseudocode}
\algnewcommand\algorithmicforeach{\textbf{Until :}}
\algnewcommand\algorithmicendif{\textbf{End}}
\algnewcommand\ForEach{\item[ \algorithmicforeach]}
\algnewcommand\EndiFF{\item[ \algorithmicendif]}

\usepackage[margin=0.63in]{geometry}

\begin{document}
%
\title{Joint Hybrid Backhaul and Access Links Design in Cloud-Radio Access Networks}

\author{
\IEEEauthorblockN{Oussama Dhifallah$^{\dagger}$, Hayssam Dahrouj$^{\ddagger}$, Tareq Y.Al-Naffouri$^{\dagger}$ and Mohamed-Slim Alouini$^{\dagger}$\\}
\IEEEauthorblockA{$^{\dagger}$Computer, Electrical and Mathematical Sciences and Engineering Division\\
King Abdullah University of Science and Technology (KAUST)\\
Email: \{oussama.dhifallah, tareq.alnaffouri, slim.alouini\}@kaust.edu.sa \\
$^{\ddagger}$ Department of Electrical and Computer Engineering, Effat University\\
Email: hayssam.dahrouj@gmail.com }
}
\maketitle

\begin{abstract}
The cloud-radio access network (CRAN) is expected to be the core network architecture for next generation mobile radio systems. In this paper, we consider the downlink of a CRAN formed of one central processor (the cloud) and several base-station (BS), where each BS is connected to the cloud via either a wireless or capacity-limited wireline backhaul link. The paper addresses the joint design of the hybrid backhaul links (i.e., designing the wireline and wireless backhaul connections from the cloud to the BSs) and the access links (i.e., determining the sparse beamforming solution from the BSs to the users). The paper formulates the hybrid backhaul and access link design problem by minimizing the total network power consumption. The paper solves the problem using a two-stage heuristic algorithm. At one stage, the sparse beamforming solution is found using a weighted mixed $\ell_1/\ell_2$ norm minimization approach; the correlation matrix of the quantization noise of the wireline backhaul links is computed using the classical rate-distortion theory. At the second stage, the transmit powers of the wireless backhaul links are found by solving a power minimization problem subject to quality-of-service constraints, based on the principle of conservation of rate by utilizing the rates found in the first stage. Simulation results suggest that the performance of the proposed algorithm approaches the global optimum solution, especially at high signal-to-interference-plus-noise ratio (SINR).
\end{abstract}
\IEEEpeerreviewmaketitle

\section{Introduction}
Cloud-radio access network (CRAN) technology is expected to support the tremendous requirements in mobile data traffic for next generation mobile radio systems (5G)\cite{IEEEhowto:cran}. In CRANs, base-stations (BSs) from different tiers are connected to the central processor (CP) via high capacity backhaul links. The CP then handles all processing of the baseband signals. Such centralized processing provides a powerful tool to jointly manage the interference, increase network capacity and improve energy efficiency. Such performance improvement is especially dependent on the joint provisioning of resources between the backhaul links and the heterogeneous radio access network, which especially depends on the type of backhaul connection available between the cloud and each BS. While the optical fiber backhaul is suitable for medium-to-large cells \cite{IEEEhowto:bl}, it remains an expensive backhaul solution especially in dense networks with hundreds of available base-stations. Optical fiber may also be unavailable at the required geographical location. Wireless backhauls, on the other hand, provide a cheap and scalable solution for small cell deployment \cite{IEEEhowto:hdg}; however, its performance is inferior to the fiber solution and depends on the characteristics of the wireless medium. As next generation networks are expected to be diverse in cell sizes and radio access technologies, hybrid connections between the central cloud and the base-stations are also needed, i.e., co-existence of both wire and wireless backhaul links \cite{IEEEhowto:hn}.

We consider a downlink CRAN, where each BS is connected to the cloud with either an optical fiber (wireline) or a wireless backhaul link. Each BS communicates with a set of users via wireless access links. We assume that the wireless backhauk links and the wireless access links are out-of-band, i.e., no interlink interference between the wireless backhauks and wireless access links. The network performance becomes a function of the nominal capacity of the wireline connection, the nature of the wireless backhaul connection, and the access link between the BSs and the served users. As the cloud performs a joint precoding of user signals to be transmitted over the finite-capacity wireline backhaul links, the performance becomes related to the compression scheme needed to forward the precoded signals to the corresponding BSs. Such compression induces quantization noise at each wireline link, and so determining the correlation matrix of the quantization noise becomes crucial. The performance of the wireless backhaul link, on the other hand, is a function of the wireless interference medium and of the optimized transmit power of the cloud's wireless terminal. Further, the radio-access link (BSs to users) depends on which BSs serve each user as well as the corresponding beamforming solution of each user's active set of serving BSs (also known as the group sparsity beamforming solution).

This paper addresses the above joint provisioning of resources between the wireline/wireless backhauls and the access links. The paper is in part related to the group-sparsity beamforming problem studied in  \cite{IEEEhowto:bl}, which proposes an iterative algorithm based on a weighted mixed $\ell_1/\ell_2$ norm minimization, an approach adopted from the compressed sensing literature. However, reference \cite{IEEEhowto:bl} assumes that the transport backhaul links have high-capacity, and so quantization noise from compression is simply neglected. This paper is further related to the joint precoding and compression problems studied in \cite{IEEEhowto:pc}, where a Majorization Minimization (MM) based-algorithm is proposed to solve the weighted sum-rate maximization problem. The approach considered in \cite{IEEEhowto:pc}, however, does not optimize the sparse beamforming solution of the access link problem.

Unlike previous works which consider either wireline or (exclusive) wireless backhaul links, this paper considers a hybrid wireless/wireline backhaul system. It focuses on the problem of minimizing the total network power consumption to determine the correlation matrix of the quantization noise of the wireline backhaul links, the transmit power of the wireless backhaul links, and the sparse beamforming of each user across the network. The main contribution of this paper is a two-stage heuristic iterative algorithm. At the first stage, the network power consumption minimization problem is formulated as a joint BS selection and beamforming optimization problem that can be solved using a weighted mixed $\ell_1/\ell_2$ norm. The quantization noise levels are then computed using the classical rate-distortion relationship. At the second stage, the transmit powers of the wireless backhaul links are found by fixing the solutions found in the first stage, and then reformulating the problem as a power minimization problem subject to rate constraints based on the principle of conservation of rate between the inflow (cloud to BS) and outflow (BS to users). Simulation results show the performance improvement of the proposed algorithm as compared to methods from the classical literature.

\section{System Model and Problem Formulation}
\label{sec:un}
\begin{figure}[t]
\centering
\includegraphics[width=8cm]{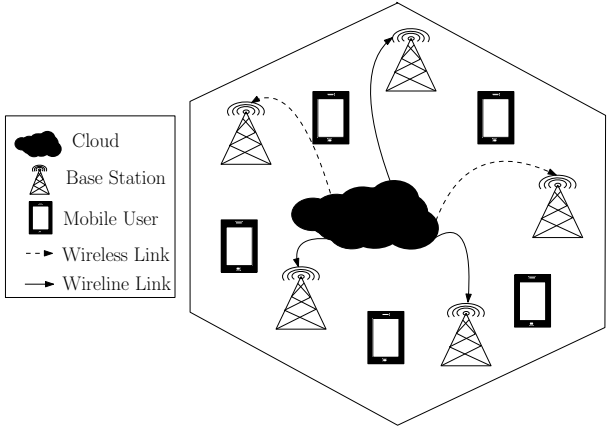}
\centering
\caption{An example of CRAN, in which, the BSs are connected to a cloud through wireline and wireless backhaul links.}
\label{mov_targ}
\end{figure}

Consider a downlink CRAN with a set of BSs connected to the cloud via $B^{wl}$ wireline capacity-limited backhaul links, and $B^{wll}$ wireless backhaul links. Let $C_l$ be the capacity limit of each wireline backhaul link $l$. Let $P_l$ be the transmit power of the $l^{th}$ wireless backhaul link. The network comprises K single-antenna Mobile Users (MUs). For simplicity of analysis, the paper assumes that both BSs and MUs are each equipped with single antenna. Figure \ref{mov_targ} illustrates a typical example of the considered model with 5 Base Stations, 3 wireline capacity-limited backhaul links, 2 wireless backhaul links and 5 MUs.

Let $\mathcal{B}=\lbrace 1, \cdots,B \rbrace$ be the set of base-stations, $\mathcal{B}^{wl}=\lbrace 1, \cdots,B^{wl} \rbrace$ the set of BSs connected to the cloud using wireline links, and  $\mathcal{B}^{wll}=\lbrace B^{wl}+1, \cdots,B \rbrace$ be the set of BSs connected to the cloud using wireless links. Further, let $\mathcal{A} \subset \mathcal{B}$ denote the set of active BSs, $\mathcal{A}^{wl} \subset \mathcal{B}$ denote the set of active BSs connected to the cloud using wireline links, and $\mathcal{A}^{wll} \subset \mathcal{B}$ denote the set of active BSs connected to the cloud using wireless links. We assume that the wireless backhauk links and the wireless access links are out-of-band, and so there is no interference between the wireless backhauks and wireless access links. The received signal $y_k \in \mathbb{C}$ at the $k^{th}$ user can be written as
\begin{equation}
	\begin{aligned}
y_{k}&=\sum\limits_{l\in \mathcal{A}}^{}~h_{lk}^{*}w_{lk}s_k+ \sum\limits_{i\ne k}^{}\sum\limits_{l\in \mathcal{A}}^{}~h_{lk}^{*}w_{li}s_i+\sum\limits_{l\in \mathcal{A}^{wl}}^{}~h_{lk}^{*}e_l+z_k,
	\label{re_wav}
\end{aligned}
\end{equation}
where $s_k$ denotes the data symbol for the $k^{th}$ user and $w_{lk} \in \mathbb{C}$ is the beamforming scalar at the $l^{th}$ BS for the $k^{th}$ user, $h_{lk}  \in \mathbb{C}$ denotes the channel scalar from the $l^{th}$ BS to the $k^{th}$ user, $\bf{e}$ $=[e_1,e_2,\cdots,e_{B^{wl}}]$ is the quantization noise assumed to be non-uniform white Gaussian process and independent of $x_{l}$ with diagonal covariance matrix $\bf{Q}$ $\in \mathbb{C}^{B^{wl}\times B^{wl}}$ with diagonal entries $q^2_l$ and $\bf{z}$ $=[z_{1},\cdots,z_K] \sim  \mathcal{CN}(0,\sigma^2\bf{I})$ is the additive Gaussian noise. Without loss of generality, we assume that $E(|s_k|^2 )= 1$ and the $s_k$’s are independent from each other. The signal-to-interference-plus-noise Ratio ($\rm{SINR}$) of user $k$ can then be expressed as

\begin{equation}
\small
	\begin{aligned}
	\rm{SINR}_k=\frac{\left| \sum\limits_{l\in \mathcal{A}}^{}~h_{lk}^{*}w_{lk} \right|^2}{\sum\limits_{i\ne k}^{}\left|\sum\limits_{l\in \mathcal{A}}^{}~h_{lk}^{*}w_{li}\right|^2+\sum\limits_{l\in \mathcal{A}^{wl}}^{}~\left|h_{lk}^{*}q_l\right|^2+\sigma^2}.
	\label{sinr}
\end{aligned}
\end{equation}

Let ${\bf w}_k=[w_{lk}]^T \in \mathbb{C}^{|\mathcal{A}|}$ and ${\bf h}_k=[h_{lk}]\in \mathbb{C}^{|\mathcal{A}|}$ be the respective beamforming and channel vectors of user $k$ due to the set of active BSs. Further, let $\tilde{\bf h}_{k}=[h_{lk}]\in \mathbb{C}^{|\mathcal{A}^{wl}|}$ be the channel vector of user $k$ due to the set of active BSs that are connected to the cloud using wireline connections. Let $\tilde{\bf Q}$ be the correlation matrix of the quantization noise of the \textit{active} wireline connections, i.e., $\tilde{\bf Q}=\textnormal{diag}(|q_l|^2)\in \mathbb{R}^{|\mathcal{A}^{wl}|\times |\mathcal{A}^{wl}|}$ indexed by $l \in\mathcal{A}^{wl}$. Therefore, we get ${\bf h}_k^H{\bf w}_k=\sum\limits_{l\in \mathcal{A}}^{}~h_{lk}^{*}w_{lk}$ and $\tilde{\bf h}_{k}^H \tilde{\bf Q} \tilde{\bf h}_{k}=\sum\limits_{l\in \mathcal{A}^{wl}}^{}~\left|h_{lk}^{*}q_l\right|^2$. The SINR expression (\ref{sinr}) can then be rewritten as
\begin{equation}
	\begin{aligned}
	\rm{SINR}_k=\frac{|{\bf h}^H_k{\bf w}_k|^2}{\sum\limits_{i\ne k}^{}|{\bf h}^H_k{\bf w}_i|^2+|\tilde{\bf h}_{k}^H \tilde{\bf Q} \tilde{\bf h}_{k}|+\sigma^2}.
	\label{re_wav}
\end{aligned}
\end{equation}
Using the rate distortion theory and assuming an independent quantization at each BS, the quantization noise level $q^2_l$, the transmit power $P_l$ of the $l^{th}$ BS connected to the cloud using wireline backhaul link, and the backhaul capacity $C_l$ are related as follows
\begin{equation}
\label{rate_distortion1}
\textnormal{log}_2\left(1+\frac{\sum\limits_{k=1}^{K}~|w_{lk}|^2}{q_l^2}\right)\leq C_l , 
\end{equation}
\begin{equation}
\label{rate_distortion2}
\sum\limits_{k=1}^{K}~|w_{lk}|^2+q^2_l\leq P_l,
\end{equation}
Furthermore, the transmission setup from the cloud to the BSs connected through the wireless links behaves as a wireless broadcast channel. The received power at the $l^{th}$ BS wirelessly connected to the cloud can be written as $P_l=|\hat{h}_l|^2\tilde{P}_l+\sum\limits_{m\ne l}^{}|\hat{h}_l|^2\tilde{P}_m+\kappa^2$, where $\hat{h}_l \in \mathbb{C}$ is the channel scalar from the cloud to the BSs connected to the cloud using wireless links, and where $\kappa^2$ is the variance of the additive Gaussian noise of the wireless backhaul. The power constraint at the $l^{th}$ BS, connected to the cloud using wireless links, can be written as
\begin{equation}
	\begin{aligned}
	\sum\limits_{k=1}^{K}~|w_{lk}|^2\leq P_l,~~\forall~~l\in \mathcal{A}^{wll}.
	\label{pow_cons}
\end{aligned}
\end{equation}

This paper considers the problem of minimizing the total network consumption which consists of the transmit power consumption of the active BSs and the relative backhaul link power consumption:
\begin{equation}
	\begin{aligned}
	&p(\mathcal{A},{\bf w})=\sum\limits_{l\in \mathcal{A}^{wl}}^{}~\left\lbrace\frac{1}{\xi_l}\left(\sum\limits_{k=1}^{K}~|w_{lk}|^2+q^2_l\right)+P^c_l\right\rbrace\\
	&~~~~~~~~~~+\sum\limits_{l\in \mathcal{A}^{wll}}^{}~\left\lbrace\frac{1}{\xi_l}\sum\limits_{k=1}^{K}~| w_{lk}|^2+P^c_l\right\rbrace ,
	\label{npc}
\end{aligned}
\end{equation}
where $\xi_l$ is the drain efficiency of the radio frequency (RF) power amplifier and $P_l^c$ denotes the relative backhaul link power consumption \cite{IEEEhowto:bl}, i.e.
\begin{equation}
\begin{cases}P_l^c=(P_{a,l}^{bs}+P_{a,l}^{onu})-(P_{s,l}^{bs}+P_{s,l}^{onu}), & \mbox{if } l\in\mathcal{A}^{wl} \\
P_l^c=P_{a,l}^{bs}-P_{s,l}^{bs}, & \mbox{if } l\in\mathcal{A}^{wll} \end{cases},
\end{equation}
where $P_{a,l}^{bs}$ and $P_{s,l}^{bs}$ denote the power consumed by the $l^{th}$ BS in the active mode and sleep mode, respectively, and where $P_{a,l}^{onu}$ and $P_{s,l}^{onu}$ denote the power consumed by the $l^{th}$ optical network units (ONU) in the active mode and sleep mode, respectively.

The paper then focuses on solving the following optimization problem
\begin{equation}
\begin{aligned}
~~\underset{{\bf w},\mathcal{A},{\bf Q}, {\tilde{P}_l}}{\operatorname{min}}&~~~~~p(\mathcal{A},{\bf w})\\	
&s.t.~\rm{SINR}_k=\frac{|{\bf h}_k^H{\bf w}_k|^2}{\sum\limits_{i\ne k}^{}|{\bf h}_k^H{\bf w}_i|^2+|\tilde{\bf h}_{k}^H \tilde{\bf Q} \tilde{\bf h}_{k}|+\sigma^2}\geq \delta_k\\
&~~~~~~\textnormal{log}_2\left(1+\frac{\sum\limits_{k=1}^{K}~|w_{lk}|^2}{q_l^2}\right)\leq C_l,~\forall~l\in\mathcal{A}^{wl}\\
&~~~~~~\sum\limits_{k=1}^{K}~|w_{lk}|^2+q^2_l\leq P_l,~\forall~l\in\mathcal{A}^{wl}\\
&~~~~~~\sum\limits_{k=1}^{K}~|w_{lk}|^2 \leq P_l,~\forall~l\in\mathcal{A}^{wll},
\end{aligned}
\label{fir_form}
\end{equation}
where the optimization is over the beamformers $\bf w$, the active set of BSs $\mathcal{A}$, the transmit power of the wireless backhauls ${\tilde{P}_l}$, and the correlation of the quantization matrix $\bf Q$, and where ${\bf \delta} =(\delta_1,\delta_2,\cdots,\delta_K)$ represents the target SINRs. The above optimization problem is of high complexity, as it involves  mixed discrete and continuous optimization problem. This paper presents a heuristic solution to solve this problem using techniques from compressed sensing and optimization theory.

\section{Joint Design of Hybrid Backhaul and Access Link}
\label{sec:deux}
In this section, a two-stage heuristic algorithm is proposed to address the complex optimization problem (\ref{fir_form}). Specifically, start with fixing the transmit power of the wireless backhaul links and solve a weighted mixed $\ell_1/\ell_2$ norm minimization problem to induce group sparsity and determine the active set of BS's and the corresponding beamforming vectors. The quantization noise levels can then be computed using the classical rate-distortion relationship. Finally, the powers of the wireless links are determined using the principle of conservation of data rate. This approach transforms the power optimization problem to a classical power minimization problem subject to quality-of-service constraints.

\subsection{Group Sparsity Formulation}
By fixing the power of the wireless backhaul links, we express the group sparsity problem by first relating the quantization noise to the beamforming vectors. This is achieved by using (\ref{rate_distortion1}) and heuristically replacing the inequality by equality, i.e.  
\begin{equation}
\begin{aligned}
  q_l^2=\frac{\sum\limits_{k=1}^{K}~|w_{lk}|^2}{2^{C_l}-1}.
\end{aligned}
\label{eqc}
\end{equation}

Note that turning the first inequality in (\ref{rate_distortion1}) into an equality in (\ref{eqc}) may not be the optimal solution to problem (\ref{fir_form}). This approach is solely used as a heuristic to make the complicated problem (\ref{fir_form}) more tractable. Typically, the inequality (\ref{rate_distortion1}) should be scaled by its corresponding Lagrangian dual variable, which only adds more complication to the problem. However, as the simulation section in this paper suggests, such heuristic approach already shows a good performance improvement as compared to classical strategies in the literature.

Based on (\ref{eqc}), we can easily show that the network power consumption minimization problem can be reformulated as
\begin{equation}
\begin{aligned}
\underset{{\bf w},\mathcal{A} }{\operatorname{min}}&~p({\bf w},\mathcal{A})=\sum\limits_{l\in\mathcal{A}}^{}\sum\limits_{k=1}^{K}~\beta_l|w_{lk}|^2+\sum\limits_{l\in\mathcal{A}}^{}~P^c_l\\
&s.t.~\rm{SINR}_k=\frac{|{\bf h}_k^H{\bf w}_k|^2}{\sum\limits_{i\ne k}^{}|{\bf h}_k^H{\bf w}_i|^2+\sum\limits_{k^{\prime}=1}^{K}|{\bf w}_{k^{\prime}}^H{\bf H}_{k}{\bf w}_{k^{\prime}}|+\sigma^2}\geq \delta_k\\
&~~~~~~\sum\limits_{k=1}^{K}~|w_{lk}|^2 \leq \hat{P}_l,~\forall~l\in\mathcal{A},
\end{aligned}
\label{sec_form}
\end{equation}
where the optimization is over ${\bf w}$ and $\mathcal{A}$, and where ${\bf H}_{k}=\frac{1}{2^{C_l}-1}\textnormal{diag}(|h_{lk}|^2,0,\cdots,0)\in \mathbb{R}^{|\mathcal{A}|\times |\mathcal{A}|}$ indexed by $l \in\mathcal{A}^{wl}$ such that ${\bf w}_{k^{\prime}}^H{\bf H}_{k}{\bf w}_{k^{\prime}}=\sum\limits_{l\in\mathcal{A}^{wl}}^{}\frac{1}{2^{C_l}-1}|h_{lk}w_{lk^{\prime}}|^2$,
\begin{equation}
\hat{P}_l=\begin{cases} \frac{2^{C_l}-1}{2^{C_l}}P_l, & \mbox{if } l\in\mathcal{A}^{wl} \\
P_l, & \mbox{if } l\in\mathcal{A}^{wll} \end{cases},
\end{equation}
and
\begin{equation}
\label{beta_def}
\beta_l=\begin{cases} \frac{2^{C_l}}{\xi_l(2^{C_l}-1)}, & \mbox{if } l\in\mathcal{A}^{wl} \\
\frac{1}{\xi_l}, & \mbox{if } l\in\mathcal{A}^{wll} \end{cases}.
\end{equation}

One can show that the optimization problem (\ref{sec_form}) can be recast as a second-order cone programming (SOCP), which can be solved efficiently, e.g., using the interior point method \cite{IEEEhowto:lp}. However, obtaining the globally optimal solution to the above minimization problem may be difficult from a computational point of view. In order to reduce the complexity, we use a similar approach to \cite{IEEEhowto:bl} and exploit the group sparsity structure of the beamforming vector ${\bf w}=[w_{11},\cdots,w_{1K},\cdots,w_{B1},\cdots,w_{BK}]^T$ where the $l^{th}$ group ${\tilde{\bf w}_l}=[w_{l1},\cdots,w_{lK}]^T$ is set to zero when the $l^{th}$ BS is switched off. Thus, the network power minimization problem (\ref{fir_form}) can be reformulated as
\begin{equation}
\begin{aligned}
\underset{{\bf w} }{\operatorname{min}}&~p({\bf w})=\sum\limits_{l=1}^{B}\sum\limits_{k=1}^{K}~\beta_l|w_{lk}|^2+\sum\limits_{l=1}^{B}~P^c_l I(\mathcal{T}({\bf w})\cap \mathcal{G}_l \ne 0)\\
&s.t.~~~C_1(\mathcal{B}),C_2(\mathcal{B}),
\end{aligned}
\label{third_form}
\end{equation}
where the optimization is over the beamformers $\bf w$, $\mathcal{T}({\bf w})=\lbrace i|{\bf w}_l(i)\ne 0\rbrace$ represents the support of the beamforming vector ${\bf w}$, $I$ represents the indicator function, and  $\mathcal{G}_l=\lbrace K(l-1)+1,\cdots,Kl\rbrace$ denotes the $l^{th}$ partition of $\mathcal{V}=\lbrace 1,\cdots,KB\rbrace$. It is easily seen that applying a phase rotation to the beamforming vectors ${\bf w}_k$ does not modify the objective function and constraints of (\ref{fir_form}). Thus, the constraints in (\ref{third_form}) can be written as (SOC) constraints, i.e.

\begin{eqnarray}
&C_1(\mathcal{B}):&\sqrt{\sum\limits_{i\ne k}^{}|{\bf h}_k^H{\bf w}_i|^2+\sum\limits_{k^{\prime}=1}^{K}|{\bf w}_{k^{\prime}}^H{\bf H}_k{\bf w}_{k^{\prime}}|+\sigma^2}\nonumber\\
& &\leq \frac{1}{\sqrt{\delta_k}} \mathcal{R}({\bf h}_k^H{\bf w}_k) \\
&C_2(\mathcal{B}):&\sqrt{\sum\limits_{k=1}^{K}~|w_{lk}|^2} \leq \sqrt{\hat{P}_l}.
\end{eqnarray}
where $\mathcal{R}(.)$ denotes the real part of the complex number.

Since the sparse reformation of the optimization problem is still computationally hard, we will follow the same approach proposed  in \cite{IEEEhowto:bl} where a tightest convex lower bound for the objective function in (\ref{third_form}) is first provided. Then, the Majorization-Minimization (MM) \cite{IEEEhowto:mm} algorithm is used in order to further induce the group sparsity. Finally, the iterative Group Sparsity Beamforming (GSBF) algorithm is used to solve the network power minimization problem (\ref{sec_form}).

Once the sparse beamforming vectors and the active BS are determined, the entries of correlation matrix of the quantization noise are found using relationship (\ref{eqc}).

\subsection{Wireless Backhaul Power Optimization}
Fixing the sparse beamforming and quantization matrix in the first stage, the transmit powers of the wireless backhaul links $\tilde{P}_l~,~l\in\mathcal{A}^{wll}$, are found based on the principle of conservation of rate between the inflow (cloud to BS) and outflow (BS to users). Accordingly, a service rate requirement (and equivalently an SINR requirement) is generated at each BS connected to the cloud through a wireless link. The transmit power optimization problem then boils down to a classical power minimization subject to quality-of-service constraints
\begin{equation}
\begin{aligned}
\underset{ \lbrace \tilde{P}_l\rbrace }{\operatorname{min}}&~p(\mathcal{A}^{wll})=\sum\limits_{l\in\mathcal{A}^{wll}}^{}\tilde{P}_l\\
&s.t.~\rm{SINR}_l=\frac{\tilde{P}_l|\hat{h}_l|^2}{\sum\limits_{m\ne l}^{}\tilde{P}_m|\hat{h}_l|^2+\kappa^2}\geq \gamma_l,~\forall l \in \mathcal{A}^{wll},
\end{aligned}
\label{LP}
\end{equation}
where the optimization is over the power $\tilde{P}_l$, and where the SINR thresholds ($\gamma_l,~l \in \mathcal{A}^{wll}$) are found using the principle of conservation of data rate. The achievable rate for the $l^{th}$ BS is first found using the parameters found in the first stage
\begin{equation}
\begin{aligned}
R_l=\sum\limits_{k=1}^{K}\textnormal{log}\left(1+\frac{|h_{lk}w_{lk}|^2}{\sum\limits_{m\ne k}^{}|h_{lk}w_{lm}|^2+\sigma^2}\right),
\end{aligned}
\label{rc}
\end{equation}
the SINR thresholds ($\gamma_l,~l \in \mathcal{A}^{wll}$) are then computed as follows
\begin{equation}
\begin{aligned}
\gamma_l=2^{R_l}-1, \forall~l~\in~\mathcal{A}^{wll}.
\end{aligned}
\label{rc}
\end{equation}

The power minimization problem (\ref{LP}) is a classical power optimization problem that can be recast as a linear program (LP), and can be solved efficiently using the interior point method \cite{IEEEhowto:lp}.
\subsection{Iterative Algorithm}
The proposed solution to solve problem (\ref{fir_form}) eventually requires to iterate between the first and the second stage. The proposed iterative group sparsity beamforming and power optimization algorithm (I-GSBPO) is summarized in Table (\ref{alg:hgsb}).
\begin{algorithm} [htb]
 \caption{The iterative group sparsity beamforming and power optimization algorithm (I-GSBPO)}
  \label{alg:hgsb}
  \begin{algorithmic}[1]
   \Require
   Initialize the transmit power $\lbrace P_{1},\cdots,P_B\rbrace$.
   \Ensure
    \State Solve the network power consumption minimization problem (\ref{sec_form}) using the the group sparse beamforming algorithm. If it is infeasible, go to {\bf End};
    \label{code:fram:extract}
   \State Compute the quantization noise levels using (\ref{eqc});
   \label{code:fram:trainbase}
    \State Solve the linear programming formulation of the power minimization problem (\ref{LP}) based on the principle of conservation of data rate. If it is infeasible, go to {\bf End};
    \ForEach The difference between the optimal network power consumption (\ref{npc}) obtained in two consecutive iterations is very small.\\
   Compute the optimal quantization noise levels, transmit power consumption of the wireless links and the beamforming vectors.
   \EndiFF
  \end{algorithmic}
\end{algorithm}

\section{Simulation results}
\label{sec:trois}
In this section, we provide some simulation results to show the performance of the proposed I-GSBPO algorithm. To this end, we consider the following channel model
\begin{equation}
\begin{cases}
h_{lk}=D_{lk}g_{lk} \\
\tilde{h}_{lk}=\tilde{D}_{lk}\tilde{g}_{lk}
 \end{cases},
\end{equation}
where $D_{lk}$ and $\tilde{D}_{lk}$ denote the large-scale fading coefficients and $g_{lk}\sim \mathcal{CN}(0,1)$,  and $\tilde{g}_{lk}\sim \mathcal{CN}(0,1)$ represent the small-scale fading coefficients.

Several methods in the literature can be used to solve the first step of the proposed heuristic algorithm. The Coordinated Beamforming (CB) algorithm \cite{IEEEhowto:hd} which only minimizes the total transmit power consumption and also assumes that all the BSs are active, the Sparsity Pattern (SP) algorithm \cite{IEEEhowto:sp} which adopts the unweighted mixed $\ell_1/\ell_2$ norm to induce group sparsity and the Greedy Selection (GS) algorithm \cite{IEEEhowto:bl} are used to measure the performance of the proposed algorithm.


\begin{figure}[h!]
\centering
\includegraphics[width=8cm]{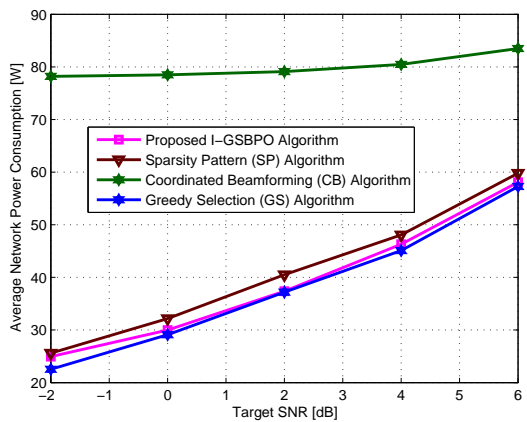}
\centering
\caption{Average Network Power Consumption versus target SINR.}
\label{anpcf}
\end{figure}

Consider a network consisting of $B=12$ single-antenna BSs and $K=4$ single-antenna MUs. Furthermore, we assume that $D_{lk}=0.051$ when $l \in \mathcal{S}_1$, $D_{lk}=0.041$ when $l \in \mathcal{S}_2$ and $D_{lk}=0.032$ when $l \in \mathcal{S}_3$ where $|\mathcal{S}_1|=|\mathcal{S}_2|=|\mathcal{S}_3|=4$, $\mathcal{S}_1\cup \mathcal{S}_2\cup \mathcal{S}_3=\mathcal{B}$ and $\mathcal{S}_i$ is uniformly drawn from $\mathcal{B}$. Furthermore, the I-GSBPO algorithm is initialized using the following transmit powers $P_l=1W$, $\forall~l \in \mathcal{B}^{wl}$ and $\tilde{P}_l=1W$, $\forall~l \in \mathcal{B}^{wll}$. We set  $\tilde{D}_{lk}=0.042$, $C_l=140$, $\xi_l=0.25$, $\forall~l \in \mathcal{B}$ and $P_l^c=(4.2+l) W$, $\forall~l \in \mathcal{B}^{wl}$ and $P_l^c=(l-5.9) W$, $\forall~l \in \mathcal{B}^{wll}$. Besides, we suppose that $\delta=0.05$, $\sigma=0.01$ and $\epsilon=\frac{0.001}{L}$.

To show the performance of the proposed algorithm as a function of multiple realizations of the channels, figure \ref{anpcf} illustrates the average network power consumption versus different SINR targets. Each point is averaged over 70 randomly and independently generated network realizations. it can be noticed that the proposed algorithm outperforms the SP and CB algorithms. Furthermore, the proposed I-GSBPO algorithm approaches global optimum solution especially at high SINR.
\begin{figure}[h!]
\centering
\includegraphics[width=8cm]{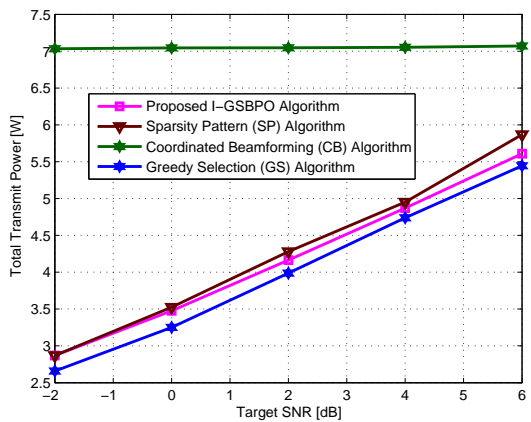}
\centering
\caption{Total transmit power versus target SINR.}
\label{tpcf}
\end{figure}

Finally, figure \ref{tpcf} shows the total transmit power consumption $p(\mathcal{A})=\sum\limits_{l\in\mathcal{A}}^{} {P}_l$ versus different SINR targets. This figure proves that both the GS algorithm and the I-GSBPO algorithm provides better performance than the SP and CB algorithms in minimizing the total transmit power consumption.

\section{Conclusion}
\label{sec:quatre}
Optimization in hybrid backhaul networks is expected to be an active area of research for next generation wireless systems. This paper considers a futuristic downlink cloud-radio access, where each BS is connected to the cloud with either a optical fiber (wireline) or a wireless backhaul link. The network performance becomes a function on the nominal capacity of the wireline connection, the nature of the wireless backhauk connection, and the access link between the BSs and the served users. The paper proposes a heuristic solution to  the joint design of the hybrid backhaul links (i.e., designing the wireline and wireless backhaul connections from cloud to BSs) and the access links (i.e., determining the sparse beamforming solution from the BSs to the users). Simulation results show the performance improvement of the proposed algorithm as compared to methods from classical literature, and that the proposed algorithm approaches the global optimum solution especially at high SINR.

\ifCLASSOPTIONcaptionsoff
  \newpage
\fi

\end{document}